\newif \ifArxiv
\Arxivtrue

\ifArxiv
\documentclass[runningheads]{llncs}
\else
\documentclass{llncs}
\fi

\usepackage{graphicx}
\usepackage{xspace}
\usepackage{amsmath}
\usepackage{color}
\usepackage[colorlinks=true]{hyperref}
\usepackage{subfigure}

\ifpdf
\hypersetup{
  pdftitle={A Tipping Point for Planarity},
  pdfauthor={E. Balloni, G. Di Battista, and M. Patrignani}
}
\fi
\hypersetup{colorlinks,linkcolor={blue},citecolor={blue},urlcolor={blue}} 

\let\doendproof\endproof
\renewcommand\endproof{~\hfill\qed\doendproof}
\newcommand{\interval}[0]{significant interval\xspace}

\begin{document}
\title{A Tipping Point for the Planarity of Small and Medium Sized Graphs%\\
%{\textcolor{red}{(This is still a draft)}}
\thanks{This research was supported in part 
        by MIUR Project ``MODE'' under PRIN 20157EFM5C, 
        by MIUR Project ``AHeAD'' under PRIN 20174LF3T8, 
        and by Roma Tre University Azione 4 Project ``GeoView''.
        \ifArxiv
Appears in the Proceedings of the 28th International Symposium on Graph Drawing and Network Visualization (GD 2020).
\else 
\fi
        }}
%
%\titlerunning{Abbreviated paper title}
% If the paper title is too long for the running head, you can set
% an abbreviated paper title here
%
%\author{Emanuele Balloni\inst{1}\orcidID{0000-1111-2222-3333} \and
%Maurizio Patrignani\inst{1}\orcidID{1111-2222-3333-4444}}
\ifArxiv
\author{%
Emanuele Balloni \and 
Giuseppe Di Battista \and 
Maurizio Patrignani}
\authorrunning{E. Balloni et al.}

\else
\author{%
Emanuele Balloni \and 
Giuseppe Di Battista\orcidID{0000-0003-4224-1550} \and 
Maurizio Patrignani\orcidID{0000-0001-9806-7411}}
\fi
%
% First names are abbreviated in the running head.
% If there are more than two authors, 'et al.' is used.
%
\institute{Roma Tre University, Rome, Italy\\ 
\email{ema-bal93@hotmail.it}\\
\email{\{giuseppe.dibattista,maurizio.patrignani\}@uniroma3.it}}
\maketitle              % typeset the header of the contribution
\begin{abstract}
This paper presents an empirical study of the relationship between the density of small-medium sized random graphs and their planarity.
It is well known that, when the number of vertices tends to infinite, there is a sharp transition between planarity and non-planarity for edge density $d=0.5$.
However, this asymptotic property does not clarify what happens for graphs of reduced size.
We show that an unexpectedly sharp transition is also exhibited by small and medium sized graphs.
Also, we show that the same ``tipping point'' behavior can be observed for some restrictions or relaxations of planarity (we considered outerplanarity and near-planarity, respectively). 
%
%It is known that when the density increases there is a sharp transition between planarity and non-planarity if the number of vertices tends to infinite.
 % moderated size? bounded size?
%We show that this ``tipping point'' behavior can be observed also in small and medium sized graphs where the transition is unexpectedly sharp. %
%We show that the same holds for some restrictions or relaxations of planarity (we considered outerplanarity and near-planarity, respectively). 
% We show that those graphs exhibit a sharp transition between  (tipping point behavior).
% This was already known for graphs whose number of vertices tends to infinity. increasing the density of a s
% The result of our investigation is somehow unexpected: observing random graphs, 
% We found a tipping point in density and size before which planarity can be given for granted and beyond which planarity is very unlikely. 
%By contrast, other graph-theoretic properties, like acyclicity, exhibit much slower transitions. 
%This paper is devoted to investigating what is the relationship between the property of being planar and the density of a graph. The result of our investigation is somehow unexpected: observing random graphs, we found that there is a tipping point in density and size before which planarity can be given for granted and beyond which planarity is very unlikely. The same holds for some restrictions or relaxations of planarity (we considered outerplanarity and near-planarity, respectively). By contrast, other graph-theoretic properties, like acyclicity, exhibit much slower transitions.
\end{abstract}

%%%%%%
%%%%%%
%%%%%%
%%%%%%
\section{Introduction}\label{se:intro}

Several popular Graph Drawing algorithms devised to draw graphs of small-medium size assume that the graph to be drawn is planar both in the static setting~\cite{t-hdg-63,fpp-hdpgg-90,s-epgg-90} and in the dynamic one~\cite{10.1007/s00454-018-0018-9,ddfpr-upm-tcs-20,bdfp-gssa-j-20}. Hence, to assess the practical applicability of such algorithms it is crucial to study the probability that a small-medium sized graph (say of about $100$--$200$ vertices) is planar. 
In particular, it is interesting to consider how this probability varies as a function of the density of the graph. We might have that the probability of planarity changes smoothly or that it changes abruptly, exhibiting a tipping-point behaviour.
%
% Thus, in this paper we study the probability that such graphs are planar. Of course, this type of study can be done from several perspectives. We adopt a pragmatic point of view, generating random graphs and performing planarity tests. Also, we focus on how planarity is related to another typical Graph Drawing parameter, the density of the graph.

A \emph{tipping point} is a threshold that, when exceeded, leads to a sharp change in the state of a system. In sociology, for example, a tipping point is a time when most of the members of a group suddenly change their behavior by adopting a practice that before was considered rare. In climate study, a tipping point is a quick and irreversible change in the climate, triggered by some specific cause, like the growth of the global mean surface temperature. Even in graph theory, tipping points have been found.
As an example, in 1960 Erd{\"o}s and R{\`e}nyi established that a random graph $G(n,m)$ with $n$ vertices and $m$ edges undergoes an abrupt change when the average vertex degree is equal to one, that is when $m \approx n/2$~\cite{er-erg-60}.
Namely, when $m = cn/2$ and $c < 1$, asymptotically almost surely the connected components are all of size $O(\log n)$, and are either trees or unicyclic graphs. Conversely, when $c > 1$, almost surely there is a unique giant component of size~$\Theta(n)$. The density $d=m/n=1/2$ is sometimes referred to as the \emph{critical density} or \emph{phase transition density}. See~\cite{b-rg-85,jlr-rg-00} for a discussion of these concepts.

In this paper we 
%seek to understand 
investigate
whether the density plays a similar role for the planarity of small-medium sized graphs. Namely, when the the density of such graphs increases, does the probability of planarity change smoothly or abruptly?

\begin{figure}[tb]
	\centering
	\subfigure[]{\includegraphics[trim=0 -30 0 0,clip,width=0.23\columnwidth]{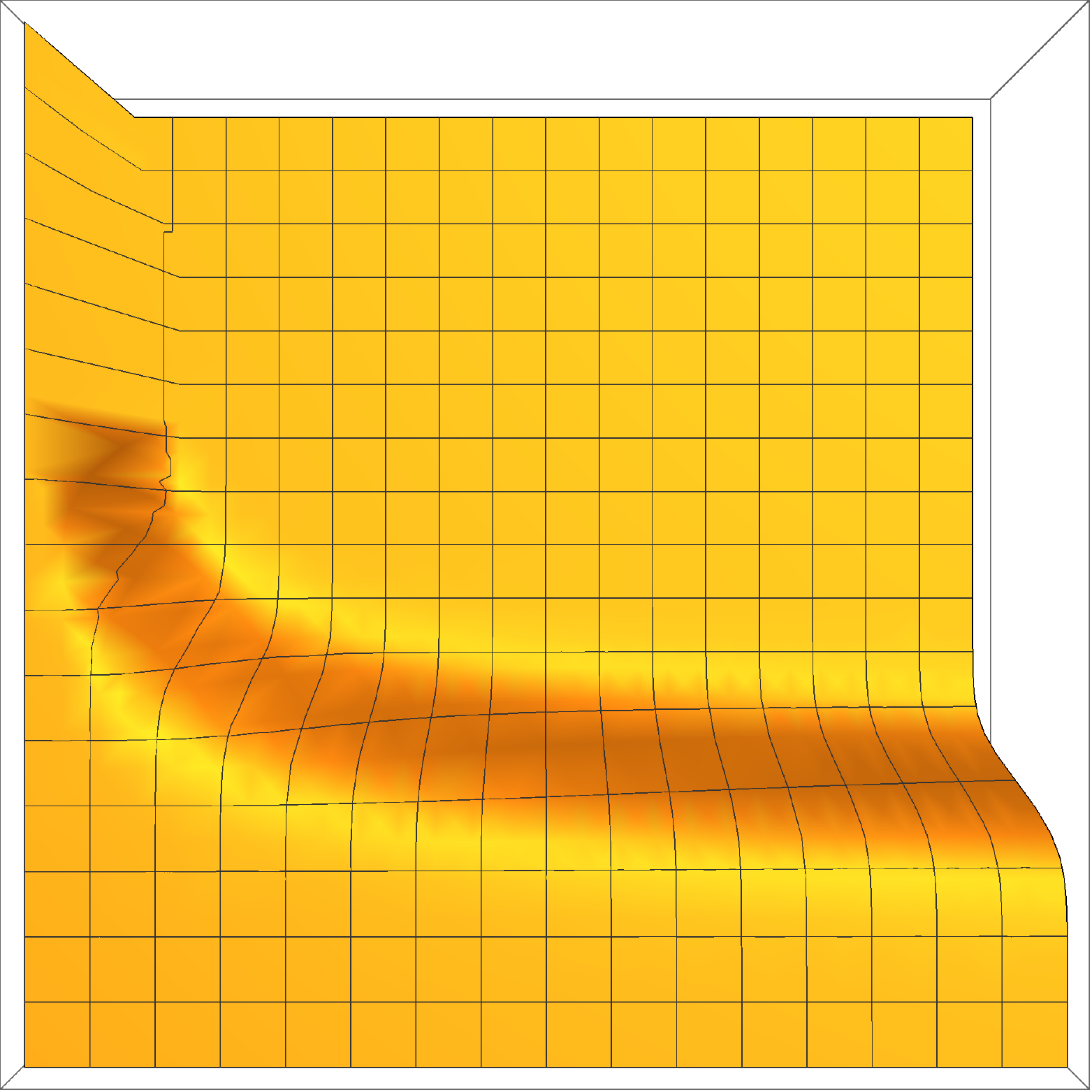}}
	\label{fig:function1}
    \begin{picture}(0,0)
	\put(-78,0){\scriptsize Number of vertices}
    \put(-92,35){\hbox{\rotatebox{90}{\scriptsize Density}}}
	\end{picture}
	\hfil
	\subfigure[]{\includegraphics[trim=0 -30 0 0,clip,width=0.23\columnwidth]{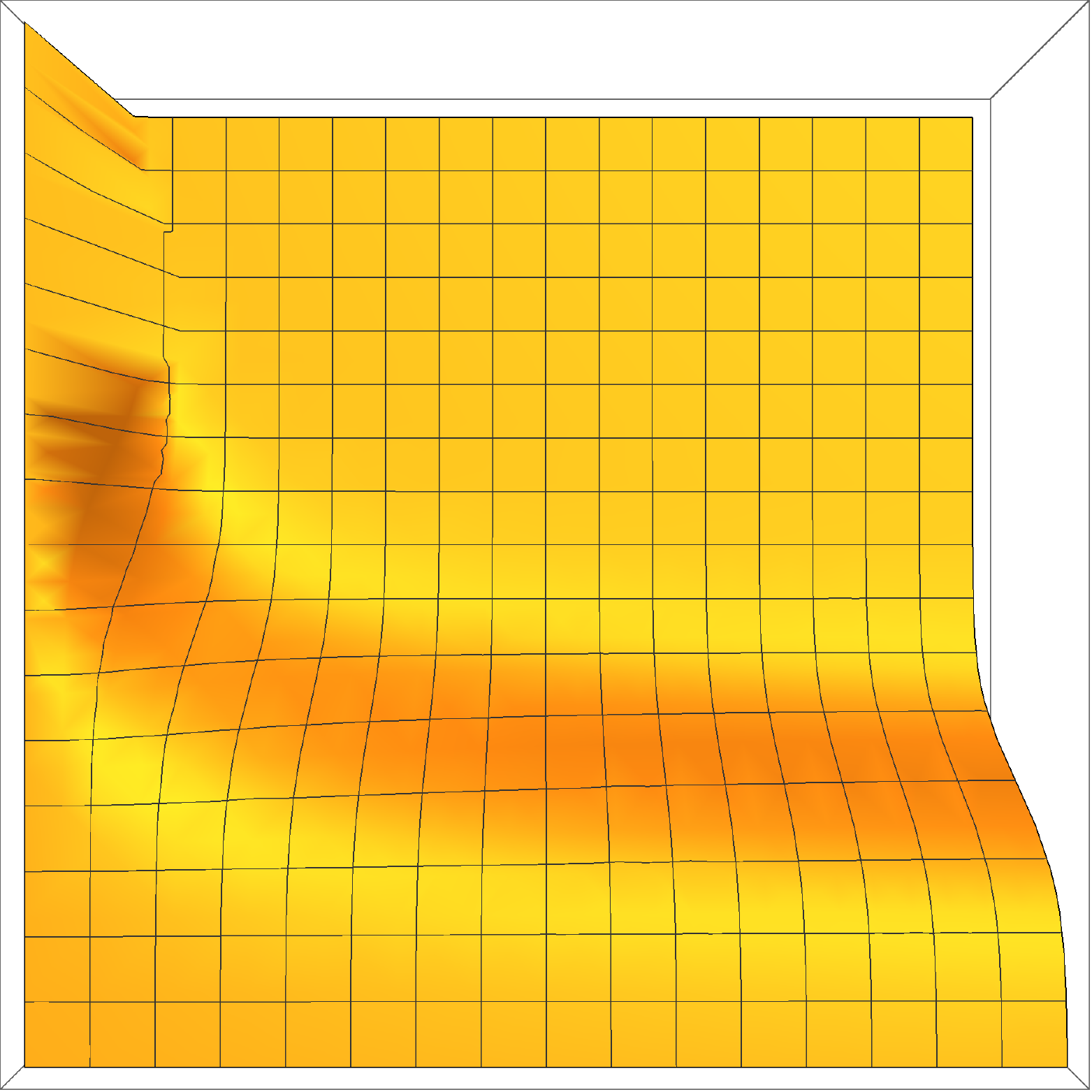}}
	\label{fig:function2}
	\begin{picture}(0,0)
	\put(-78,0){\scriptsize Number of vertices}
    %\put(-90,35){\hbox{\rotatebox{90}{\scriptsize Density}}}
	\end{picture}
	\hfil
	\subfigure[]{\includegraphics[trim=0 -30 0 0,clip,width=0.23\columnwidth]{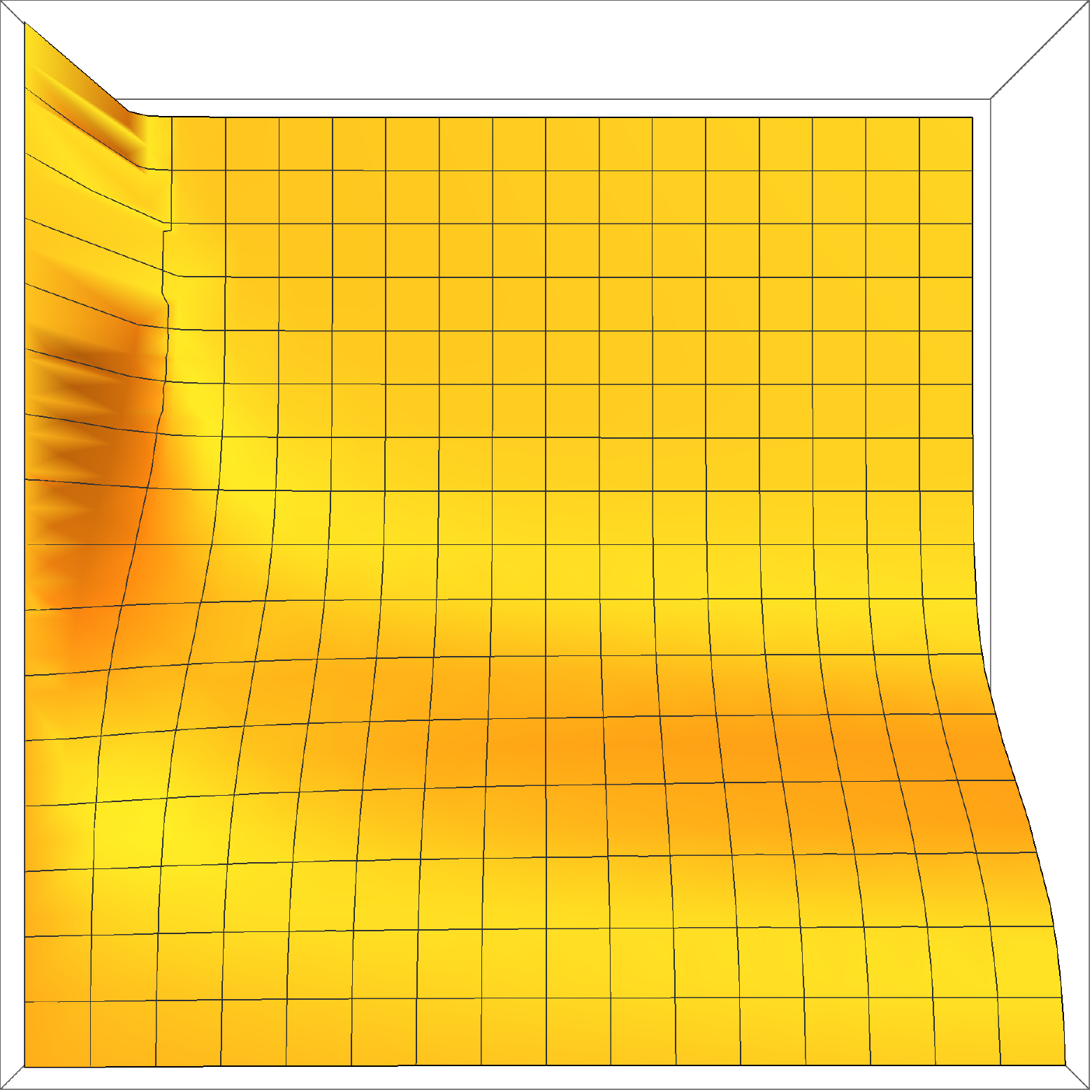}}
	\label{fig:function3}
	\begin{picture}(0,0)
	\put(-78,0){\scriptsize Number of vertices}
    %\put(-90,35){\hbox{\rotatebox{90}{\scriptsize Density}}}
	\end{picture}
	\hfil
	\subfigure[]{\includegraphics[trim=0 -30 0 0,clip,width=0.23\columnwidth]{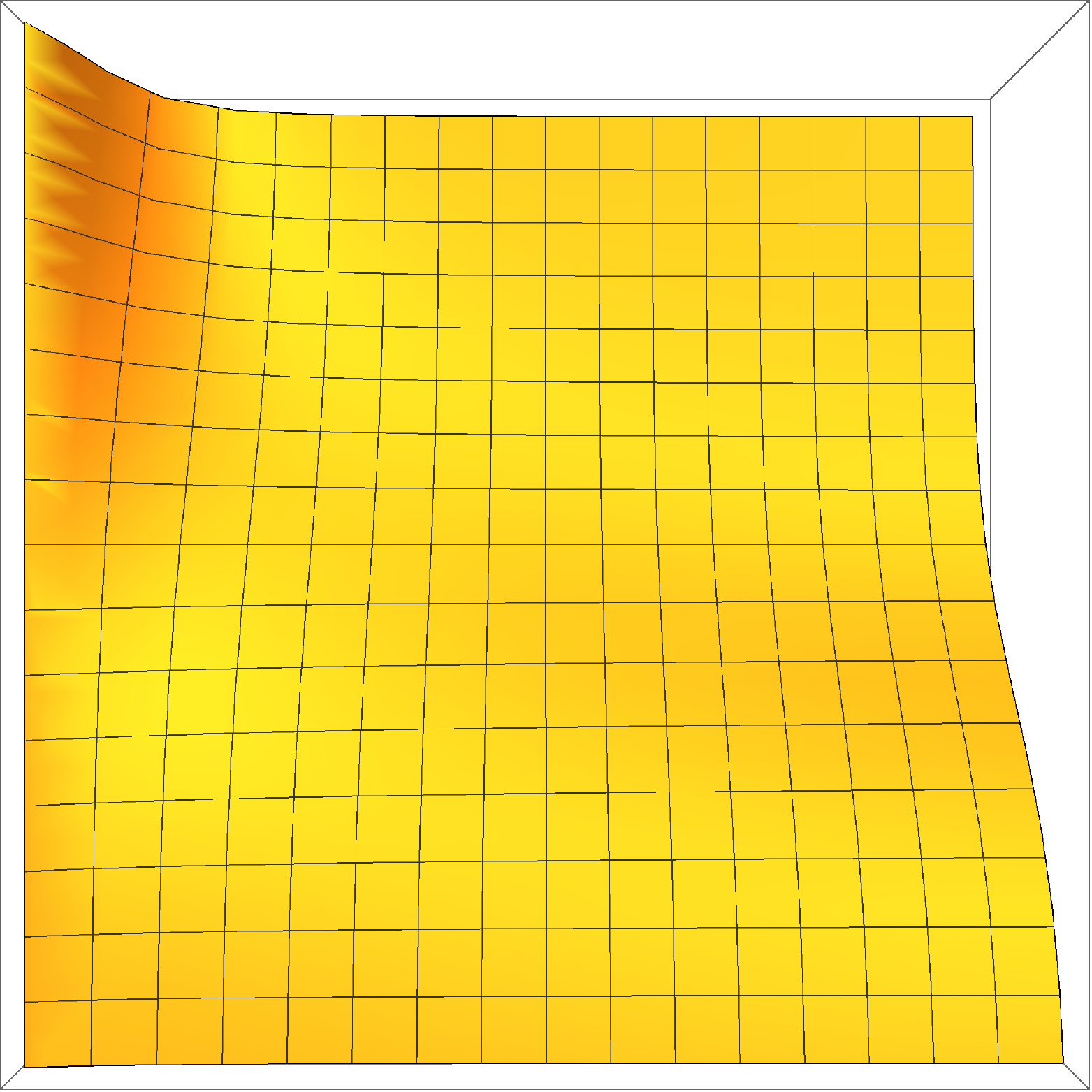}}
	\label{fig:function4}
	\begin{picture}(0,0)
	\put(-78,0){\scriptsize Number of vertices}
    %\put(-90,35){\hbox{\rotatebox{90}{\scriptsize Density}}}
	\end{picture}
	\hfil
	\caption{Function $\zeta(n,d)$ for $n \in [1,400]$ and $d \in [0,3]$ in four cases: 
	         (a) $c_1$=$5$, $c_2$=$0.5$, $c_3$=$20$, $c_4$=$0.5$; 
	         (b) $c_1$=$5$, $c_2$=$0.5$, $c_3$=$8$, $c_4$=$0.5$; 
	         (c) $c_1$=$5$, $c_2$=$0.5$, $c_3$=$4$, $c_4$=$0.5$; and 
	         (d) $c_1$=$10$, $c_2$=$0.5$, $c_3$=$1$, $c_4$=$0.5$.
	         }\label{fig:function}
\end{figure}
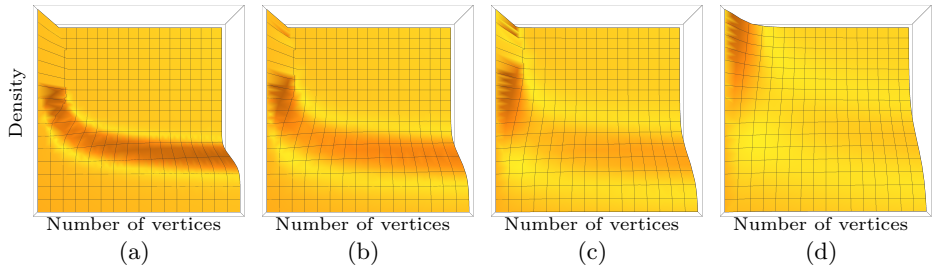

To answer this question one could think of using the result of {\L}uczak {\em et al.}~\cite{lpw-srgpp-94} who show that a random graph is almost surely non-planar if and only if the number of edges is $n/2 + O(n^{2/3})$. From the point of view of the density this means that a graph is almost surely non-planar if the density is $1/2 + O(n^{-1/3})$. However, the result shows only an asymptotic bound and does not clarify what happens for small-medium sized graphs.
Essentially, this means that, for $n \rightarrow \infty$ graphs with density greater than $1/2$ are almost surely non-planar and that the ``transition range'' of density  within which the probability of planarity falls from $1$ to $0$ is $\Theta(n^{-1/3})$. This result has been confirmed in~\cite{nrr-pprgn-13}, where it is proved that a graph with infinitely many vertices and density $1/2$ has probability $\approx0.998$ to be planar. Again, this gives no hint about how large is in practice this transition range for small values of $n$.
For example, Fig.~\ref{fig:function} shows four plots for different values of the constants $c_1,\dots,c_4$ of the function $\zeta(n,d)$ which has both the asymptotic behaviors described in~\cite{lpw-srgpp-94}
\ifArxiv
(see Appendix~\ref{app:zeta}).
\else
(see~\cite{bdp-tppsm-20-arxiv}).
\fi 

$$\zeta(n,d) = \frac{1}{2^{(d-(0.5+c_1/n^{c_2}))\cdot(c_3 + c_4 n^{1/3})}+1}$$

Depending on the values of $c_1,\dots,c_4$ the function shows quite different behaviours in the range $n \in [1,400]$.

In this paper we adopt a pragmatic point of view. Namely, we are interested into investigating what are the properties of a random graph of small-medium size $n$ when its density increases. 
%
%Our findings are coherent with the theoretical results in the literature and contribute to depict the complex scenario of random graph properties. 
%
In particular, we experimentally measured that, for each graph size $n \leq 400$, there is a value of density that marks a sharp transition from planar graphs to non-planar ones. 
% We propose approximations of the function $\delta(n)$ for planar, outerplanar, and near-planar graphs. 
This behavior is shared also by restrictions or relaxations of planarity, such as outerplanarity, and near-planarity. 

The paper is structured as follows. Section~\ref{se:methods} describes the methodology used for all experiments. Section~\ref{se:experiments} describes each experiment in detail. Our conclusions are given in Section~\ref{se:conclusions}. 

%%%%%%%
%%%%%%%
%%%%%%%
%%%%%%%
\section{Experimental Setting}\label{se:methods}

All the experiments described in Section~\ref{se:experiments} are composed of three phases: generation of graphs; measurement; and analysis. In this section we describe the characteristics of the three phases common to all experiments. 
%In Section~\ref{se:experiments} we specify additional details for each experiment. 

\smallskip\noindent
\textbf{Generation of graphs.} In all experiments (but for near-planarity) we used graphs with a number $n$ of vertices that varies from $1$ to $400$, increasing at each step by one. 
The density $d=\frac{m}{n}$, where $m$ is the number of edges, varies in a range that depends on the type of property that we are investigating. In fact, given a specific property, there always exists an interval of densities, that we call the \emph{\interval}, such that for a graph outside the \interval either the property is granted or the property is ruled out, while inside the \interval there are both graphs that have the property and graphs that do not. 
%In practice, 
This is the interval of densities that we aim to experimentally explore\footnote{For the smallest graphs we may not have all densities. For example, there is no graph with $5$ vertices and density greater than $2$.}. 

For each combination of size $n$ and density $d$ we determined the number of edges $m = \textrm{Round}(n\cdot d)$ of the graphs to be generated, and generated $10,000$ random graphs with $n$ vertices and $m$ edges\footnote{Function $\textrm{Round}()$ rounds a value to the nearest integer, where $\textrm{Round}(0.5)=1.0$.}.
In particular, we used function \texttt{randomSimpleGraph} of the OGDF library~\cite{cgjkkm-ogdf-14} for uniformly-at-random generating labeled graphs with a given number of vertices and edges. All graphs were simple (no loops or multiple edges allowed). 

\smallskip\noindent
\textbf{Measurement.} 
For each combination of size and density we counted how many graphs have the desired property. 
  
\smallskip\noindent
\textbf{Analysis.} 
%The first analysis was obviously visual, i.e., performed by plotting and inspecting the data. 
We used Wolfram Mathematica 12.0.0.0 for producing the plots that are in this paper. In particular, we used function \texttt{ListPlot3D} that joins points with flat polygons. 
For the property of acyclicity it is also possible to compute the exact percentage of random graphs that are acyclic. This allowed us to compare the measured frequency distribution with its probability counterpart
\ifArxiv
(see Appendix~\ref{app:validation}). 
\else 
(see~\cite{bdp-tppsm-20-arxiv}). 
\fi 
% TOGLIERE?
We used Mathematica also for sampling contour lines of surfaces and for computing fitting functions of sets of value pairs.

%In some other cases, by sampling contour lines and by fitting curves we were able to propose functions that describe the observed phenomena. 

%%%%%%
%%%%%%
%%%%%%
%%%%%%
\section{Experimental Results}\label{se:experiments}

In this section we report the results of the experiments to determine how density and size impact graph-theoretic properties of random graphs of small-medium size. Since the purpose of the experiments is to show that planarity exhibits a tipping point behavior when the density increases, we start our experiments with acyclicity, a property that notoriously does not have tipping points~\cite[p. 118]{b-rg-85}.
Then, we consider planarity, outerplanarity, and near-planarity, the main targets of our investigation. 
%(Sections~\ref{sse:planarity}, \ref{sse:outerplanar}, and~\ref{sse:near-planar}, respectively).
%All experiment were performed on...

%
%%%
%%%%%
%%%
%
%\subsection{Acyclicity in Random Graphs}\label{sse:acyclicity}

%In this section we investigate how density and size impact the property of random graphs of being acyclic. 

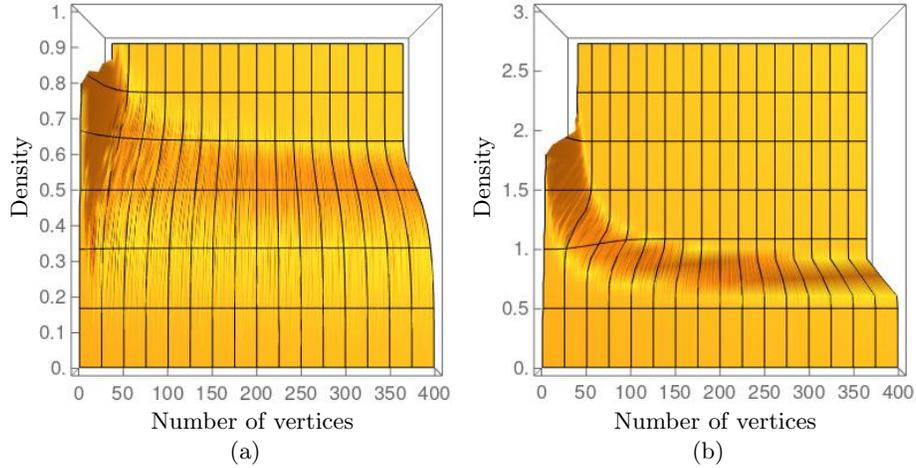
\begin{figure}[tb]
	\centering
	\hfill
	\subfigure[]{\includegraphics[trim=0 -20 0 0,clip,width=0.45\columnwidth]{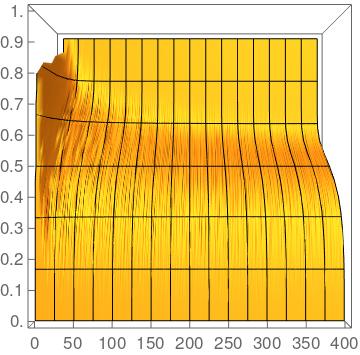}
	\label{fig:acyclic-from-above}
	}
	\begin{picture}(0,0)
	\put(-120,0){Number of vertices}
    \put(-173,80){\hbox{\rotatebox{90}{Density}}}
	\end{picture}
	\hfill
	\subfigure[]{\includegraphics[trim=0 -20 0 0,clip,width=0.45\columnwidth]{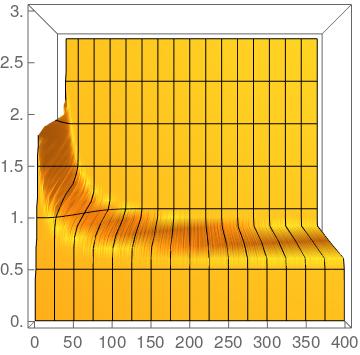}
	\label{fig:planarity-from-above}
	}
	\begin{picture}(0,0)
	\put(-120,0){Number of vertices}
    \put(-173,80){\hbox{\rotatebox{90}{Density}}}
	\end{picture}
	%\hfil
	\caption{(a) Measured fraction of random graphs that are acyclic. (b) Measured fraction of random graphs that are planar.}\label{fig:acyclic-and-planar}
\end{figure}

\smallskip\noindent\textbf{Acyclicity in Random Graphs.}
Simple graphs with less than three edges are acyclic. Conversely, since a tree has $m=n-1$ edges, when $m=n$ a graph has at least one cycle. Hence, the \interval of densities for acyclicity is $[\frac{3}{n},1-\frac{1}{n}]$. We used densities ranging from $0.0$ to $1.0$, with a step of~$0.05$ performing a total of $84 \times 10^6$ tests. 
%For each combination of number of vertices and density we generated $1,000$ random simple graphs and computed the percentage of acyclic graphs.
%
%\smallskip\noindent\textbf{Analysis.} 
The plot in Fig.~\ref{fig:acyclic-from-above} shows the measured frequency of acyclic graphs as a function of density and size. Is it apparent that the density is the main cause of the loss of acyclicity, while the size of the graph seems to have weaker effects. In particular, bigger graphs tend to loose acyclicity earlier than smaller graphs. 
%This could have been predicted: intuitively, adding edges to the graph creates a forest of trees; the bigger is the graph the more likely is that the size of these trees is not uniform and the bigger trees have higher probability of loosing acyclicity than the smaller ones. 

Overall, the percentage of acyclic graphs seems to decrease smoothly through the \interval of densities, without any quick transition or drop. 
Acyclic graphs allow us to compare a case where the tipping point is absent with the cases discussed in the next sections where a tipping point is present.
Also, for acyclicity we were able to compute the actual probability of a graph of having this property and we used the comparison between experimental and theoretical values to validate the experimental pipeline 
\ifArxiv
(see Appendix~\ref{app:validation}). 
\else 
(see~\cite{bdp-tppsm-20-arxiv}). 
\fi 

%
%%%
%%%%%
%%%
%
%\subsection{Planarity in Random Graphs}\label{sse:planarity}
\smallskip\noindent\textbf{Planarity in Random Graphs.} 
We now consider the property of the graph of being planar. 
All graphs with less than $9$ edges are planar and there is no planar graph with more than $3n-6$ edges. Hence, the \interval of densities for planarity is $[\frac{9}{n},\frac{3n-6}{n}]$. For our experiments we used densities from $0.0$ to $3.0$, with a step of~$0.1$, performing a total of $124 \times 10^6$ planarity tests.
In order to test the generated graphs for planarity we first used the OGDF function \texttt{makeConnected} that adds the minimum number of edges to make the graph connected and then called a single planarity test on the obtained graph: it can be easily seen that the minimality of the added edges implies that the connected graph is planar if and only if the connected components of the original graph were all planar. 

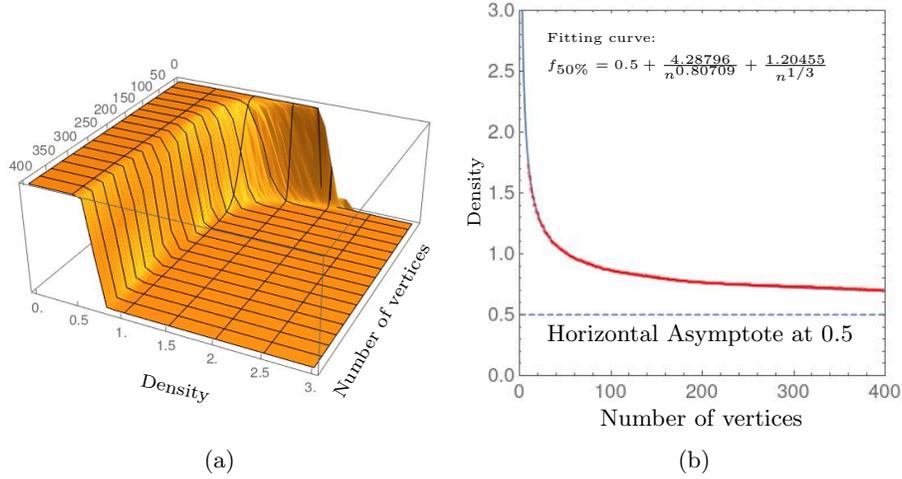
\begin{figure}[tbp]
	\centering
	\subfigure[]{\includegraphics[trim=0 -50 0 0,width=0.48\columnwidth]{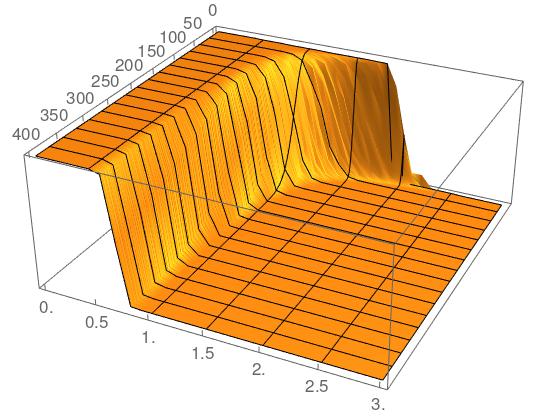}
	\label{fig:planarity-from-the-side}
	}
	\begin{picture}(0,0)
	    \put(-120,20){\hbox{\rotatebox{-13}{\scriptsize Density}}}
        \put(-47,15){\hbox{\rotatebox{60}{\scriptsize Number of vertices}}}
	\end{picture}
	\hfill\hfill
	\subfigure[]{\includegraphics[trim=0 -30 0 0,clip,width=0.45\columnwidth]{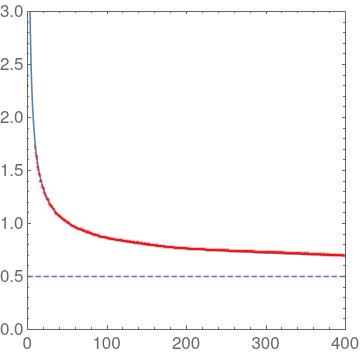}
	\label{fig:planarity-fitting}
	}
	\begin{picture}(0,0)
	\put(-120,5){Number of vertices}
    \put(-170,80){\hbox{\rotatebox{90}{\scriptsize Density}}}
    %c1 -> 4.28796, c2 -> 0.80709, c3 -> 1.20455
    \put(-140,150){\tiny Fitting curve: }
    \put(-140,140){\tiny $f_{50\%}= 0.5 + \frac{4.28796}{n^{0.80709}} + \frac{1.20455}{n^{1/3}}$}
    \put(-140,37){\footnotesize Horizontal Asymptote at $0.5$}
        \end{picture}
	\caption{(a) View from the side of the same graph of Fig.~\ref{fig:planarity-from-above}. (b) The samples at height $50\%$ (red dots) and a possible fitting curve (solid blue line).}\label{fi:planarity}
\end{figure}

Figs.~\ref{fig:planarity-from-above} and~\ref{fig:planarity-from-the-side} show a plot of the frequency of planar graphs in random simple graphs as a function of density and size.
It is apparent that the percentage of planar graphs drops from $100\%$ to $0\%$ in a short range of density values. 
As an example, for $n=200$ we have that the fraction of planar graphs drops from $99\%$ to $1\%$ in the interval of densities 
% DATI PER PLANARITA' 10%-90% 
% {200.273, 0.866346}
% {200.431, 0.682526}
% the difference is 0.18382
% 0.18382 / 3 = 0.06127
% $[0.68,0.87]$, that corresponds to the $6\%$ 
%
% DATI PER PLANARITA' 1%-99%
% approssimazionelow[200] = 0.914894
% approssimazionehigh[200] = 0.597208
% delta[200] = 0.317685
% delta[200]/3.0 = 0.105895
%
$[0.915,0.598]$, that corresponds to the $10.6\%$ 
of the significant interval. In contrast, for the same value of $n$, the fraction of acyclic graphs depicted in Fig.~\ref{fig:acyclic-from-above} drops from $99\%$ to $1\%$ in
% DATI PER ACICLICITÀ' 10%-90%
% {200.387, 0.639043}
% {200.551, 0.347391}
% la differenza è = 0.291652
% the $29\%$ of the significant interval.
%
% DATI PER ACICLICITÀ' 1%-99%
% {200.602, 0.708108}
% {200.173, 0.178082}
% la differenza e' = 0.530026
the $53\%$ of the significant interval.
The tipping point is strongly related with density and appears earlier in larger graphs.
\ifArxiv
Figure~\ref{fig:planarity-contours} in the Appendix shows a plot of $9$ equally spaced contour lines at height $10\%, 20\%, \dots, 90\%$.
\else
Figure~\ref{fig:planarity-contours} in~\cite{bdp-tppsm-20-arxiv} shows a plot of $9$ equally spaced contour lines at height $10\%, 20\%, \dots, 90\%$.
\fi 

In order to quantitatively study the behavior of the plot we determined the sample points of the contour line at height $50\%$ and computed a fitting of such points. For the fitting, because of the results in \cite{lpw-srgpp-94}, we selected a function of type $d=1/2+c_1/n^{c_2}+c_3/n^{1/3}$. The result of the fitting is shown in Fig. \ref{fig:planarity-fitting}. Observe that the value of $c_2$ is consistent with the theory.

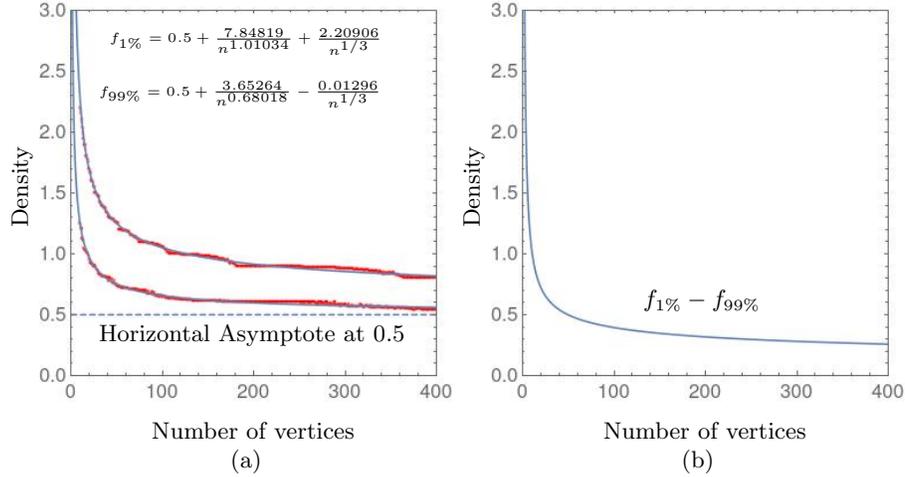
\begin{figure}[tbp]
	\centering
	\subfigure[]{\includegraphics[trim=0 -30 0 0,clip,width=0.45\columnwidth]{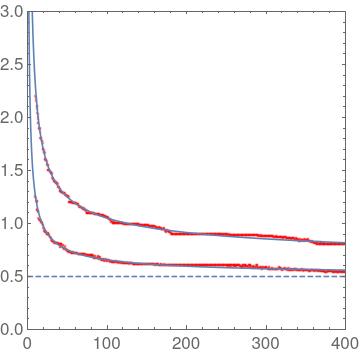}
	\label{fig:planarity-fitting-low-high}
	}
	\begin{picture}(0,0)
	\put(-120,0){Number of vertices}
    \put(-173,80){\hbox{\rotatebox{90}{Density}}}
% 99% = {c1 -> 3.65264, c2 -> 0.68018, c3 -> -0.0129579}
% 1% = {c1 -> 7.84819, c2 -> 1.01034, c3 -> 2.20906}
    %\put(-110,135){\tiny Fitting curve: }
    \put(-136,150){\tiny $f_{1\%}=0.5 + \frac{7.84819}{n^{1.01034}} + \frac{2.20906}{n^{1/3}}$}
    \put(-140,130){\tiny $f_{99\%}=0.5 + \frac{3.65264}{n^{0.68018}} - \frac{0.01296}{n^{1/3}}$}
    \put(-140,37){\footnotesize Horizontal Asymptote at $0.5$}
    \end{picture}
	\hfil
	\subfigure[]{\includegraphics[trim=0 -30 0 0,clip,width=0.45\columnwidth]{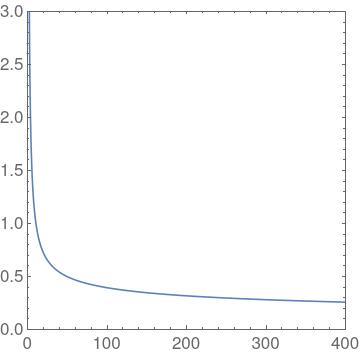}
	\label{fig:planarity-delta}
	}
	\begin{picture}(0,0)
	\put(-120,0){Number of vertices}
    \put(-173,80){\hbox{\rotatebox{90}{Density}}}
    \put(-105,50){\footnotesize $f_{1\%} - f_{99\%}$}
    \end{picture}
	\hfil
	\caption{(a) The sample points of the contour lines at height $1\%$ and $99\%$ and the corresponding fitting curves. (b) Difference between the fitting curves in (a).}\label{fig:planarity-delta-pair}
\end{figure}

In order to evaluate the width of the transition range we determined the sample points of the contour lines at height $1\%$ and $99\%$ and computed two fittings, one for each set of such points. For both the fittings, again, we selected a function of type $d=1/2+c_1/n^{c_2}+c_3/n^{1/3}$. The result are shown in Fig.~\ref{fig:planarity-fitting-low-high}. Observe how the difference between the two curves is very small (Fig.~\ref{fig:planarity-delta}).

% fitting curve closely follows the sample points

% 50% = c1 -> 4.28796, c2 -> 0.80709, c3 -> 1.20455
% 99% = {c1 -> 3.65264, c2 -> 0.68018, c3 -> -0.0129579}
% 1% = {c1 -> 7.84819, c2 -> 1.01034, c3 -> 2.20906}

% Figure~\ref{fig:planarity-fitting} shows that the  (even around the peak that we have for small values of $n$, see Fig.~\ref{fig:planarity-ripple}).
%a graph with infinite vertices and density $0.5$ has probability $\approx0.998$ to be planar~\cite{nrr-pprgn-13}.

Surprisingly, for random graphs of small-medium size the drop value for the measured fraction of planar graph is much smaller than it would have been hoped for: if you grow the density of a random graph of small-medium size you very likely loose planarity way before you have any chance to get connectivity ($d=1$). Practically speaking, if you were interested into graphs with density one, planarity is almost granted for number of vertices in the range $[1,40]$ but is almost absent above $100$ vertices.
For density $1.5$, instead, a random graph with more than $25$ vertices is very likely non-planar. 
%has a negligible probability of being planar.

%
%%%
%%%%%
%%%
%
%\subsection{Outerplanarity in Random Graphs}\label{sse:outerplanar}
\smallskip\noindent\textbf{Outerplanarity in Random Graphs.} 
An \emph{outerplanar} graph is a graph that admits a planar drawing where all vertices are on the external face. All graphs with less than $6$ edges are outerplanar --- the smallest non-outerplanar graphs being $K_4$ and $K_{2,3}$ --- and there is no outerplanar graph with more than $2n-3$ edges. Hence, the \interval of densities for outerplanarity is $[\frac{6}{n},\frac{2n-3}{n}]$. For our experiments we used densities from $0.0$ to $2.0$, with a step of~$0.1$.
%
%In order to test outerplanarity we added a vertex to the random graph, connecting it to all other vertices. Then, a single planarity test launched on the obtained connected graph provides an outerplanarity test of the original graph. 

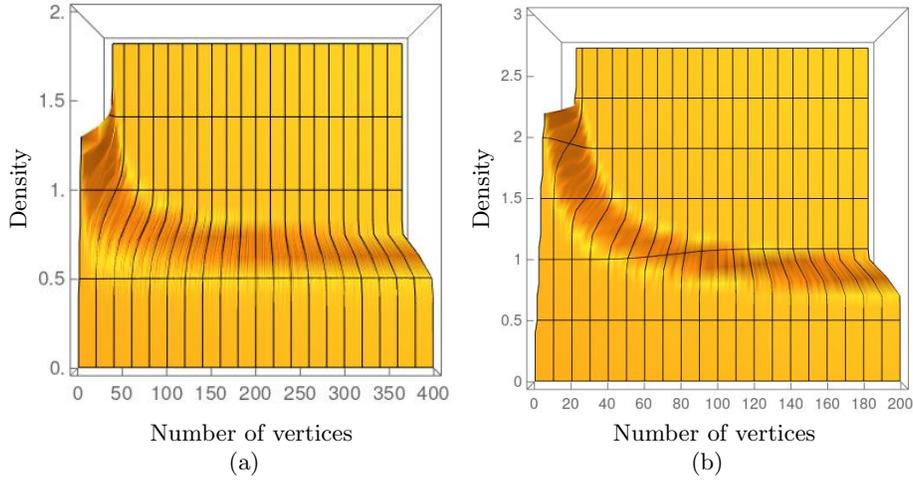
\begin{figure}[htbp]
	\centering
	\hfill
	\subfigure[]{\includegraphics[trim=0 -30 0 0,clip,width=0.45\columnwidth]{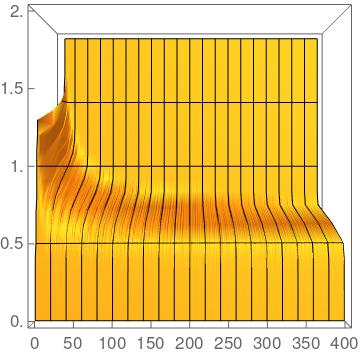}
	\label{fi:outerplanar}
	}
	\begin{picture}(0,0)
	\put(-120,0){Number of vertices}
    \put(-173,80){\hbox{\rotatebox{90}{Density}}}
	\end{picture}
	\hfill
	\subfigure[]{\includegraphics[trim=0 -30 0 0,clip,width=0.45\columnwidth]{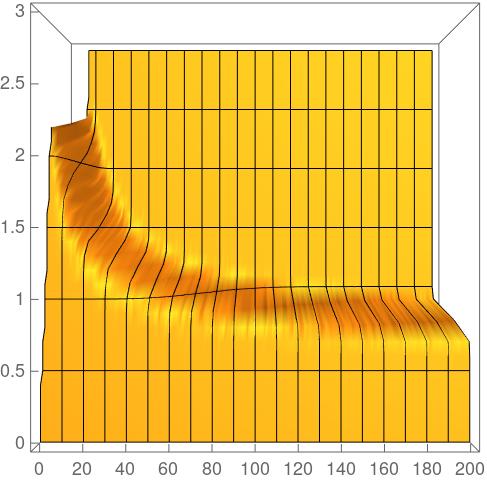}
	\label{fi:near-planar}
	}
	\begin{picture}(0,0)
	\put(-120,0){Number of vertices}
    \put(-173,80){\hbox{\rotatebox{90}{Density}}}
	\end{picture}
	\caption{(a) Measured fraction of random graphs that are outerplanar. (b) Measured fraction of random graphs that are near-planar.}\label{fi:outer-near-planar}
\end{figure}

Figure~\ref{fi:outerplanar} shows the fraction of outerplanar graphs as a function of the number of vertices and density. 
% The measured value of the \pino is $\approx 0.613$. 

%
%%%
%%%%%
%%%
%
%\subsection{Near-Planarity in Random Graphs}\label{sse:near-planar}
\smallskip\noindent\textbf{Near-Planarity in Random Graphs.}
A \emph{near-planar} graph is a graph that can be made planar by removing (at most) one edge~\cite{cm-aoepg-13}. Near-planar graphs are also called \emph{skewness-$1$} or \emph{almost planar} graphs~\cite{dlm-sgdbp-19}. The smallest not near-planar graph is $K_{3,4}$, with $12$ edges. 
From the definition of near-planar graphs it follows that such graphs have a maximum of $3n-6+1$ vertices. Hence, the \interval of densities for near-planarity is $[\frac{14}{n},\frac{3n-5}{n}]$. In our experiments we used densities ranging from $0.0$ to $3.0$ increasing by $0.1$.  
The recognition of near-planar graphs can be made in quadratic-time: it suffices to test for planarity any graph obtained by removing one edge. 

%Tempo di esecuzione: 114558 secondi, cioe' 1909.3 minuti. (31.81 ore)

Figure~\ref{fi:near-planar} shows the measured
fraction of random graphs that are near-planar as a function of the number of vertices (from $1$ to $200$) and the density.
Observe that the transition from near-planar graphs to non-near-planar ones is sharper than what we measured for planarity or quasi-planarity, although it occurs for higher values of densities. 
%In particular, the \pino value for near-planarity is $\approx 0.716$. 

%%%%%%%
%%%%%%%
%%%%%%%
%%%%%%%
%%%%%%%
\section{Conclusion and Future Work}\label{se:conclusions}

We reported empirical evidence of the existence of a tipping point for planarity in random graphs of small-medium size. 
%We propose a function $\delta(n)$, with an asymptote at $GDB \approx0.688$, to describe the density at which the abrupt drop of the percentage of planar graphs occurs. 
The same phenomenon appears to be present for restrictions and relaxations of planarity as outerplanarity and near-planarity. 
% Concerning future work, 
It would be interesting to measure whether other popular families of `beyond planar' graphs, as 1-planar or quasiplanar graphs, also feature the same abrupt transition in their distribution in random graphs. Unfortunately, testing 1-planarity is  NP-complete~\cite{km-moihp-13} even for near-planar graphs~\cite{cm-aoepg-13} and, to our knowledge, no implementation of the FPT algorithm in~\cite{bce-pc1p-18} for testing 1-planarity is available. Also, no testing algorithm has been proposed for quasi-planarity.
% Due to the limitations of the available graph generators, in our experiments we only considered labeled graphs. An experimentation with unlabeled graph would better address the purpose of our investigation. 
Finally, we could consider other types of graphs, as random bipartite, biconnected, or triconnected graph, as well as other graph models like small-world graphs or scale-free graphs. 

%
%%%
%%%%%
%%%
%
\subsection*{Acknowledgments}

%We are indebted with Giuseppe Di Battista for interesting conversations and suggestions that helped us improving our work and steering our investigation. 
We thank Carlo Batini for posing us the first question about rapid transitions of graph properties. Sometimes questions are more important than answers. We also thank the anonymous reviewer for pointing out that the smallest not near-planar graph in terms of number of edges is $K_{3,4}$.

%%%%%%%%
%%%%%%%%
%%%%%%%%
%%%%%%%%
%\clearpage
\bibliographystyle{splncs04}
\bibliography{bibliography}

\begin{thebibliography}{10}
\providecommand{\url}[1]{\texttt{#1}}
\providecommand{\urlprefix}{URL }
\providecommand{\doi}[1]{https://doi.org/#1}

\bibitem{bce-pc1p-18}
Bannister, M., Cabello, S., Eppstein, D.: Parameterized complexity of
  1-planarity. J. Graph Algorithms Appl.  \textbf{22}(1),  23--49 (2018)

\bibitem{10.1007/s00454-018-0018-9}
Barrera-Cruz, F., Haxell, P., Lubiw, A.: Morphing schnyder drawings of planar
  triangulations. Discrete Comput. Geom.  \textbf{61}(1),  161--184 (2019)

\bibitem{b-rg-85}
Bollob{\'a}s, B.: Random graph. Academic Press Inc., Harcourt Brace Jovanovich
  Publishers, London (1985)

\bibitem{bdfp-gssa-j-20}
Borrazzo, M., {Da Lozzo}, G., {Di Battista}, G., Frati, F., Patrignani, M.:
  Graph stories in small area. Journal of Graph Algorithms and Applications
  \textbf{24}(3),  269--292 (2020). \doi{10.7155/jgaa.00530}

\bibitem{cm-aoepg-13}
Cabello, S., Mohar, B.: Adding one edge to planar graphs makes crossing number
  and 1-planarity hard. SIAM Journal on Computing  \textbf{42}(5),  1803--1829
  (2013)

\bibitem{cgjkkm-ogdf-14}
Chimani, M., Gutwenger, C., J\"unger, M., Klau, G.W., Klein, K., Mutzel, P.:
  Open {G}raph {D}rawing {F}ramework ({OGDF}). In: Tamassia, R. (ed.) Handbook
  of Graph Drawing and Visualization, chap.~17. CRC Press (2014)

\bibitem{ddfpr-upm-tcs-20}
{Da Lozzo}, G., {Di Battista}, G., Frati, F., Patrignani, M., Roselli, V.:
  Upward planar morphs. Algorithmica  (2020). \doi{10.1007/s00453-020-00714-6}

\bibitem{dlm-sgdbp-19}
Didimo, W., Liotta, G., Montecchiani, F.: A survey on graph drawing beyond
  planarity. {ACM} Comput. Surv.  \textbf{52}(1),  4:1--4:37 (2019).
  \doi{10.1145/3301281}

\bibitem{er-erg-60}
Erd{\"o}os, P., R{\'e}enyi, A.: On the evolution of random graphs. Magyar Tud.
  Akad. Mat. Kutat{\'o}o Int. K{\"o}zl.  \textbf{5},  17--61 (1960)

\bibitem{fpp-hdpgg-90}
de~Fraysseix, H., Pach, J., Pollack, R.: How to draw a planar graph on a grid.
  Combinatorica  \textbf{10}(1),  41--51 (1990). \doi{10.1007/BF02122694}

\bibitem{jlr-rg-00}
Janson, S., {\L}uczak, T., Rucinski, A.: Random graphs. Wiley-Interscience
  Series in Discrete Mathematics and Optimization, Wiley-Interscience, New York
  (2000)

\bibitem{km-moihp-13}
Korzhik, V.P., Mohar, B.: Minimal obstructions for 1-immersions and hardness of
  1-planarity testing. Journal of Graph Theory  \textbf{72}(1),  30--71 (2013).
  \doi{10.1002/jgt.21630}

\bibitem{lpw-srgpp-94}
{\L}uczak, T., Pittel, B., Wierman, J.C.: The structure of a random graph at
  the point of the phase transition. Transactions of the American Mathematical
  Society  \textbf{341}(2),  721--748 (1994)

\bibitem{m-clt-70}
Moon, J.W.: Counting Labeled Trees. Canadian Mathematical Monographs, William
  Clowes and Sons (1970)

\bibitem{nrr-pprgn-13}
Noy, M., Ravelomanana, V., Ru{\'e}, J.: The probability of planarity of a
  random graph near the critical point. In: International Conference on Formal
  Power Series and Algebraic Combinatorics (FPSAC 2013). pp. 791--802 (2013)

\bibitem{s-epgg-90}
Schnyder, W.: Embedding planar graphs on the grid. In: Johnson, D.S. (ed.)
  Proceedings of the First Annual {ACM-SIAM} Symposium on Discrete Algorithms.
  pp. 138--148. {SIAM} (1990)

\bibitem{t-hdg-63}
Tutte, W.T.: {How to Draw a Graph}. Proceedings of the London Mathematical
  Society  \textbf{s3-13}(1),  743--767 (01 1963).
  \doi{10.1112/plms/s3-13.1.743}

\end{thebibliography}

\ifArxiv

%%%%%%%%
%%%%%%%%
%%%%%%%%
%%%%%%%%
\clearpage
\appendix
\section{Appendix}

%
%%%
%%%%%
%%%
%
\subsection{Asymptotic Study of Function $\zeta(n,d)$}\label{app:zeta}

% Sul piano (n,m) ho uno spazio verticale (m) che va come m = n^{-2/3}
% I nostri grafici sono sul piano (n,d)
% Sappiamo che n*d = m
% n*d = n^{-2/3}
% d = n^{-2/3}*n^{-3/3} = n^{-5/3}

Function $\zeta(n,d)$, mentioned in Section~\ref{se:intro} is defined as follows: 

\begin{equation}\label{eq:zeta}
p = \zeta(n,d) = \frac{1}{2^{(d-(0.5+c_1/n^{c_2}))\cdot(c_3 + c_4 n^{1/3})}+1}
\end{equation}

%Fig.~\ref{fig:function} shows four plots for different values of the constants $a, b, c,$ and $d$ of function $\zeta(x,y)$. 
In this section we show that: (i) the transition value for which $\zeta(n,d) = 0.5$ when $n \rightarrow \infty$ is equal to $d = 0.5$ and (ii) the transition range of numbers of edges within which the probability of planarity falls from 1 to 0 is $O(n^{2/3})$ and, hence, the transition range expressed with respect to density $d = m/n$ is $O(n^{-1/3})$.
For the first statement, consider the function $d = \psi(n,p)$ obtained by converting Equation~\ref{eq:zeta} to be explicit with respect to the density $d$:

%\begin{center}
%\textcolor{red}{(To be commented on submission)}
%\end{center}
%\textcolor{red}{
%\begin{equation*}
%p \cdot (2^{(d-(0.5+c_1/n^{c_2}))\cdot(c_3+c_4  n^{1/3})}+1) = 1 
%\end{equation*}
%\begin{equation*}
%2^{(d-(0.5+c_1/n^{c_2}))\cdot(c_3+c_4 n^{1/3})}+1 = 1/p 
%\end{equation*}
%\begin{equation*}
%2^{(d-(0.5+c_1/n^{c_2}))\cdot(c_3+c_4 n^{1/3})} = 1/p - 1 
%\end{equation*}
%\begin{equation*}
%(d-(0.5+c_1/n^{c_2}))\cdot(c_3+c_4 n^{1/3}) = \log_2(1/p - 1) 
%\end{equation*}
%\begin{equation*}
%d-(0.5+c_1/n^{c_2}) = \frac{\log_2(1/p - 1)}{c_3+c_4 %n^{1/3}}
%\end{equation*}
%}

\begin{equation}\label{eq:psi}
d = \psi(n,p) = \frac{\log_2(1/p - 1)}{c_3+c_4 n^{1/3}} + (0.5+c_1/n^{c_2})
\end{equation}

It is immediate to observe that, provided that $c_2, c_5 > 0$, the limit for $n \rightarrow \infty$ of $\psi(n,0.5)$, where $p=0.5$ is meant to capture the transition point, is $d=0.5$.

Second, we consider the transition range along the $d$-coordinate where $\zeta(n,d)$ falls from $p_{max} = 0.9$ to $p_{min} = 0.1$ (any other pair of constants $p_{max} > p_{min}$ being equivalent). This is given by $\psi(n,p_{min})-\psi(n,p_{max})$. In order to show that this range falls as $n^{-5/3}$, we divide this quantity by $n^{-5/3}$ and show that the limit for $n \rightarrow \infty$ of the obtained function is a constant. In fact we have:

%\textcolor{red}{
%\begin{equation*}
%\psi(n,p_{min}) - \psi(n,p_{max}) = \frac{\log_2(1/p_{min} - 1)-\log_2(1/p_{max} - 1)}{c_3+c_4 n^{1/3}} 
%\end{equation*}
%\begin{equation*}
%\frac{\psi(n,p_{min}) - \psi(n,p_{max})}{n^{-1/3}} = \frac{1}{n^{-1/3}}\cdot \frac{\log_2(1/p_{min} - 1)-\log_2(1/p_{max} - 1)}{c_3+c_4 n^{1/3}}
%\end{equation*}
%}

\begin{equation}\label{eq:delta}
\frac{\psi(n,p_{max}) - \psi(n,p_{min})}{n^{-1/3}} = \frac{\log_2(1/p_{min} - 1)-\log_2(1/p_{max} - 1)}{c_3 n^{-1/3} +c_4} 
\end{equation}

Since the numerator of Equation~\ref{eq:delta} is a constant and since $\lim_{n \rightarrow \infty}(n^{-1/3}) = 0$, we have that the limit for $n \rightarrow \infty$ of Equation~\ref{eq:delta} is also a constant.

%
%%%
%%%%%
%%%
%
\subsection{Comparison with Theoretical Values}\label{app:validation}

For the case of acyclicity, we were able to compute the actual probability of a random graph $G(n,m)$ to be acyclic. In fact, since the number of possible edges in a simple $n$-vertex graph is ${n}\choose{2}$, the number $g_{n,m}$ of labeled simple graphs with $m$ edges is $g_{n,m}={{{n}\choose{2}}\choose{m}}$.  
On the other hand, the number $f_{n,k}$ of labeled forests with $n$ vertices and $k$ connected components %is~\cite[page 28]{m-clt-70}:
is~\cite{m-clt-70}:
\begin{equation}\label{eq:moon}
f_{n,k} = {n \choose k} \sum_{i=0}^k {(-\frac{1}{2})^i (k+i)i!{k \choose i}{{n-k}\choose{i}}n^{n-k-i-1}}
\end{equation}

where ${x \choose y}$ is assumed to be zero when $y > x$. 
Since each edge added to a forest decreases the number of connected components by one, we have $k = n-m$ (also, recall that in a forest $m < n$). Therefore, from Equation~(\ref{eq:moon}) we can obtain the number $f_{n,m}$ of labeled forests with $m$ edges, plot the ratio $\frac{f_{n,m}}{g_{n,m}}$, and compare this function with the computed percentage. A plot of the probability $\frac{f_{n,m}}{g_{n,m}}$ of $R(n,m)$ to be acyclic for $n \in [1,400]$ and $d=m/n \in [0,1]$ is in Fig.~\ref{fig:acyclic-formula-from-above}.

Figure~\ref{fig:acyclic-comparison}, shows the absolute value of the difference between the computed frequency and the actual probability. The average of such values is $0.132\%$ with the peak value at $1.57\%$. In particular, most of the noise seem to occur in the regions where the fraction of cyclic and acyclic graphs is more balanced. If instead of the absolute value we consider the error with its sign, we have an average error of  $-0.0036\%$. This low value is an indication that the error is not biased towards higher or lower values with respect to the theoretical ones. 

Overall, for the above discussed reasons we conclude that the experimental pipeline is sound. 

\begin{figure}[tb]
	\centering
	\hfill
	\subfigure[]{\includegraphics[trim=0 -30 0 0,clip,width=0.45\columnwidth]{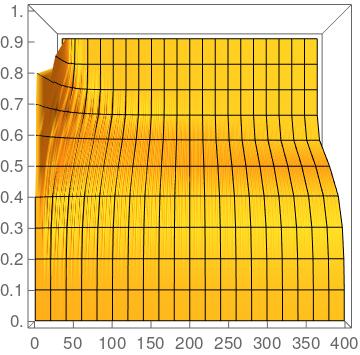}
	\label{fig:acyclic-formula-from-above}
	}
    \begin{picture}(0,0)
	\put(-120,0){Number of vertices}
    \put(-173,80){\hbox{\rotatebox{90}{Density}}}
	\end{picture}
	\hfill
	\subfigure[]{\includegraphics[trim=0 -100 0 0,width=0.44\columnwidth]{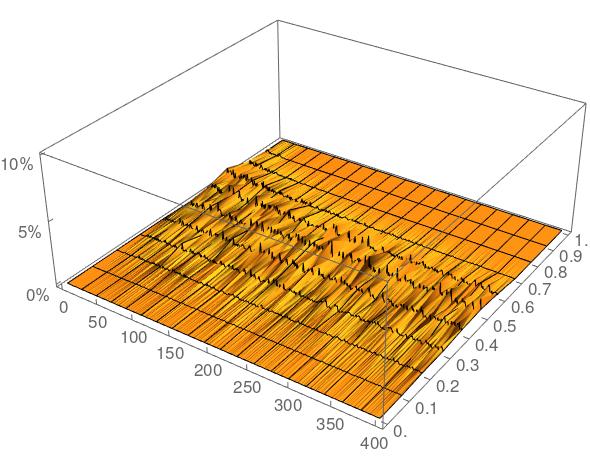}
	\label{fig:acyclic-comparison}
	}
        \begin{picture}(0,0)
	    \put(-140,49){\hbox{\rotatebox{-22}{\scriptsize Number of vertices}}}
        \put(-32,37){\hbox{\rotatebox{52}{\scriptsize Density}}}
	    \end{picture}
	\hfil
	\caption{(a) Probability of random graphs to be acyclic. (b) The absolute value of the difference between the measured frequency and the theoretical probability of the graph to be acyclic.}\label{fig:acyclic}
\end{figure}
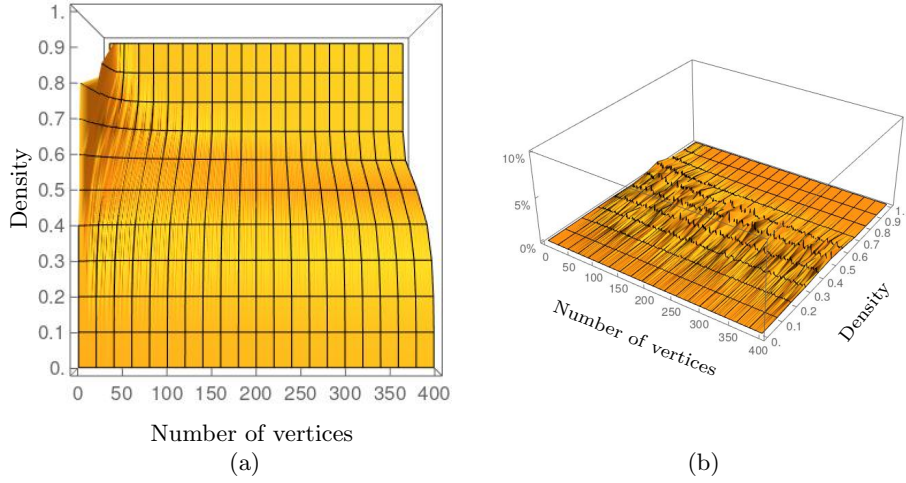

%
%%%
%%%%%
%%%
%
%\subsection{Additional Experiments}\label{app:additional}

\begin{figure}[tbp]
	\centering
	\subfigure[]{\includegraphics[trim=0 -20 0 0,clip,width=0.60\columnwidth]{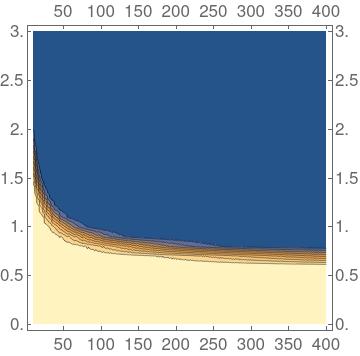}
	\label{fig:planarity-contours}
	}
	\begin{picture}(0,0)
	\put(-145,0){Number of vertices}
    \put(-225,95){\hbox{\rotatebox{90}{Density}}}
	\end{picture}
	\hfil
%	\subfigure[]{\includegraphics[trim=0 -30 0 0,clip,width=0.50\columnwidth]{figures/planarity-ripple.jpg}
%	\label{fig:planarity-ripple}
%	}
%        \begin{picture}(0,0)
%	    \put(-165,64){\hbox{\rotatebox{-50}{\scriptsize Number of vertices}}}
%        \put(-50,25){\hbox{\rotatebox{43}{\scriptsize Density}}}
%	    \end{picture}
%	\hfil
	\caption{%(a) 
	A plot of $9$ equally-spaced contour lines of the same graph of Fig.~\ref{fig:planarity-from-above} at height $10\%, 20\%, \dots, 90\%$. 
	%(b) A picture of the measured fraction of planar graph from which it is apparent the ripple for small values of $n$.
	}\label{fi:planarity-analysis}
\end{figure}
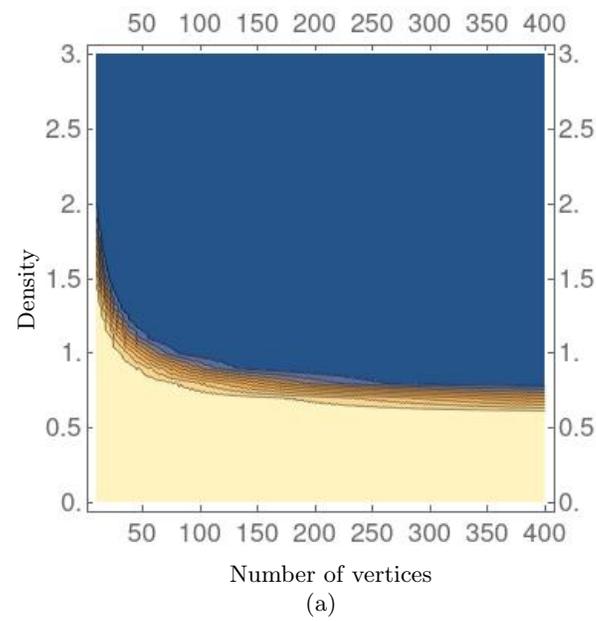

\fi  % fine dell' \ifArxiv

\end{document}